\documentclass[12pt]{article}

\usepackage{amssymb}
\usepackage{subfigure}
\usepackage{color}
\usepackage{epsfig,amssymb,amsfonts,amsmath,graphicx,dsfont,cite,xfrac}
\usepackage{authblk}
\definecolor{mygray}{gray}{0.5}

\usepackage{cite}
\usepackage[colorlinks=true,linkcolor=blue,citecolor=red]{hyperref}

\newcommand{\be}{\begin{equation}}
\newcommand{\ee}{\end{equation}}
\newcommand{\bea}{\begin{eqnarray}}
\newcommand{\eea}{\end{eqnarray}}

\title{Laguerre-Gaussian wave propagation in parabolic media}

\author[${1}$]{S. Cruz y Cruz}
\author[${2}$]{Z. Gress}
\author[${3}$]{P. Jim\'enez-Mac\'ias}
\author[${3}$]{O. Rosas-Ortiz}

\affil[${1}$]{\footnotesize Instituto Polit\'ecnico Nacional, UPIITA, Av I.P.N 2580, C.P. 07340, M\'exico City, Mexico}

\affil[${2}$]{\footnotesize Universidad Aut\'onoma del Estado de Hidalgo, Ciudad del Conocimiento, Hidalgo, Mexico}

\affil[${3}$]{\footnotesize Physics Department, Cinvestav, AP 14-740, 07000
M\'exico City, Mexico}

\date{}
\begin{document}

\maketitle

\begin{abstract}
We report a new set of Laguerre-Gaussian wave-packets that propagate with periodical self-focusing and finite beam width in weakly guiding inhomogeneous media. These wave-packets are solutions to the paraxial form of the wave equation for a medium with parabolic refractive index. The beam width is defined as a solution of the Ermakov equation associated to the harmonic oscillator, so its amplitude is modulated by the strength of the medium inhomogeneity. The conventional Laguerre-Gaussian modes, available for homogenous media, are recovered as a particular case.
\end{abstract}


\section{Introduction}

The study of optical beams having complex structures is a subject of intense activity in current times, mainly because the properties of structured light open new possibilities for the manipulation of individual atoms and small molecules \cite{Coh98,Wei99}. This subject represents a feedback pathway between theory and experiment: theoretical advances suggest new experiments while  significant experimental results require either new theoretical models or improvements in our understanding of the behavior of light. Remarkably, after realizing that azimuthally phased beams carry angular momentum \cite{All92}, it was understood that the concept of photon angular momentum is not limited to spin \cite{All00}, but it may include either extrinsic or intrinsic orbital angular momentum \cite{Ber98} (for a recent discussion on the matter see e.g. \cite{Cha18}). However, it is important to emphasize that, although spin and orbital angular momentum behave quite similar in some instances, ``orbital angular momentum has its own distinctive properties and its own distinctive optical components''  \cite{Pad00}. Such a subtlety is fundamental in the investigation of light-matter interactions \cite{Gri03,Car05,Dho06}. In this context it is notable that the wavefront structure of the Laguerre-Gaussian beams allows the production of force fields that have no counterpart in conventional optical beams \cite{Dav05,Bra05}. From the practical point of view, it has been found that Hermite-Gaussian beams with no orbital angular momentum can be transformed into Laguerre-Gaussian beams carrying orbital angular momentum \cite{Bei93,Baz90,Cru16}. Thus, one can use either cylindrical lenses \cite{Bei93} or Fork diffractive gratings \cite{Baz90,Cru16} to produce Laguerre-Gaussian beams in the laboratory (sophisticated spatial light modulators can be used  instead). Nevertheless, the propagation of Laguerre-Gaussian beams in free (homogeneous) space implies that the corresponding beam width diverges as the propagation variable increases, which may tie down the usefulness of such beams.

In this paper we address the problem of finding Laguerre-Gauss wave-packets with finite beam width along all the propagation axis. With this aim we solve the paraxial form of the wave equation for a weakly guiding inhomogeneous medium, the refractive index of which is quadratic (parabolic) in the coordinates transverse to the propagation. Our method is based on the approach introduced in \cite{Cas13}, where a Gaussian wave-packet is used to solve the Schr\"odinger equation for time-dependent and nonlinear Hamiltonian operators via complex Riccati equations. The main point in \cite{Cas13} is that the width of the packet is defined as a solution of the Ermakov equation \cite{Erm80} associated to the one-dimensional oscillator. Such approach was already applied to study the propagation of waves in non-homogeneous media \cite{Cru17,Gre17,Gre19}, were the close relationship between the paraxial wave equation and the Schr\"odinger equation is successfully exploited to construct Hermite-Gaussian wave-packets for quadratic refractive index optical media. In the present case the beam width is an oscillatory solution of the Ermakov equation that depends on the propagation variable and such that its amplitude is modulated by the strength of the medium inhomogeneity. The Laguerre-Gaussian wave-packets reported here correspond to non-dispersive beams of finite transverse optical power (localized beams) that propagate with periodical self-focusing profile in the medium. Our approach generalizes the methods to define finite beam widths already reported by other authors \cite{Tie65,Kog65,Kri80,Perm96,Bor03,Kho10}, and confirm that the distinctive angular momentum properties of the Laguerre-Gaussian modes are better prepared and exploited if the related beam propagates in parabolic media. 

The generalities to construct the above described Laguerre-Gaussian wave-packets are outlined in Section~\ref{sec2}, where it is also shown that the conventional Laguerre-Gaussian modes arise after turning-off the inhomogeneity of the medium, just as a particular case. In Section~\ref{discute} we summarize our results and provide some directions for future work.

\section{Paraxial wave equation for parabolic media}
\label{sec2}

Consider the $z$-propagation of waves through a weakly guiding inhomogeneous medium, the  refractive index of which is quadratic in the transverse coordinates (the $xy$-plane). Using polar coordinates to write the position vector transverse to beam propagation as $\boldsymbol{\rho} = (\rho, \theta)$, the refractive index we are dealing with is of the form
\begin{equation}
\label{parabolic1}
  n^2(\rho) = n_0^2 \left(1 - \Omega^2 \rho^2\right), \quad \Omega^2 \rho^2 \ll 1.
\end{equation}
Here, $n_0$ stands for the refractive index at the optical axis and $\Omega \geq 0 $ is a parameter that characterizes the focusing properties of the medium. The corresponding paraxial wave equation is given by
\begin{equation}
\label{phe1}
-\frac1{2 k_0^2 n_0} \nabla_{\perp}^2  U + \frac{n_0}2\Omega^2 \rho^2 U = \frac i{k_0} \frac{\partial U}{\partial z}.
\end{equation}
The solutions $U = U(\boldsymbol{\rho},z)$  of (\ref{phe1}) describe the transversal amplitude of the electric field in the medium. Hereafter $\nabla_{\perp}^2$ stands for the transversal component of the Laplacian operator  and $k_0$ is the wave number in free space. 

\subsection{Lowest-order Gaussian mode}
\label{lowest}

Following \cite{Cas13,Cru17,Gre17}, as a fundamental nonstationary solution of the paraxial wave equation (\ref{phe1}), we propose the Gaussian wave-packet 
\begin{equation}
\label{gwp1}
U(\rho,z) = N(z) e^{iS(z)\rho^2}.
\end{equation}
The straightforward calculation shows that the normalization factor $N(z)$ and the coefficient $S(z)$ are respectively given by
\begin{equation}
N(z) = \frac{N_0}{w(z)} e^{-i \chi(z)},  \quad S(z) = \frac{k_0n_0}2 \frac d{dz} \ln w(z)+ \frac i{w^2(z)},
\end{equation}
where $N_0$ is a normalization constant, $w(z)$ is a solution of the Ermakov equation for the one-dimensional harmonic oscillator
\begin{equation}
\label{Ermakov}
\frac{d^2w}{dz^2} + \Omega^2 w = \frac{4}{k_0^2n_0^2 w^3},
\end{equation}
and
\begin{equation}
\chi(z) = \frac2{k_0n_0} \int^z \frac{1}{ w^2(x)} dx.
\end{equation}
It is a matter of substitution to verify that after the identification
\begin{equation}
\frac1{R(z)} = \frac d{dz} \ln w(z), 
\label{radio}
\end{equation}
we can rewrite Eq.~(\ref{gwp1}) in the `canonical' form of the lowest-order Gaussian mode
\begin{equation}
\label{LG00}
U(\rho,z) = \frac{N_0}{w(z)} \exp \left( \frac{- \rho^2}{w^2(z)} \right)
\exp \left( i \left[\frac{k_0n_0\rho^2}{2R(z)}  - \chi(z)\right]\right).
\end{equation}
In analogy with the homogenous case \cite{Sie86,Sal91}, we see that $N_0$ refers to the maximum electric field strength, $w(z)$ corresponds to the beam width, $R(z)$ to the radius of curvature, and $\chi(z)$ to the Gouy phase. Notice however that the main difference between the wave-packet (\ref{LG00}) and the Gaussian mode of the homogeneous case \cite{Sie86,Sal91} is the $z$-dependence of the beam width. Indeed, we have already mentioned that $w(z)$ in (\ref{LG00}) is a solution of the Ermakov equation (\ref{Ermakov}),  so we follow  \cite{Ros15} to get
\begin{equation}
\label{width1}
w(z) = w_0\left[\cos^2 \left[\Omega (z-z_0)\right] + \frac1{\left(\Omega z_R\right)^2} \sin^2 \left[\Omega (z-z_0)\right]\right]^{1/2},
\end{equation}
with $z_0$ an integration constant, $w_0 = w(z_0)$, and $z_R = \tfrac12 k_0n_0w_0^2$. 

\begin{figure}[htb]

\centering
\subfigure[$w(z)$]{\includegraphics[width=0.3\textwidth]{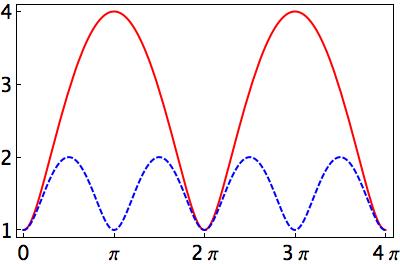} } 
\hskip1cm
\subfigure[$R(z)$]{\includegraphics[width=0.3\textwidth]{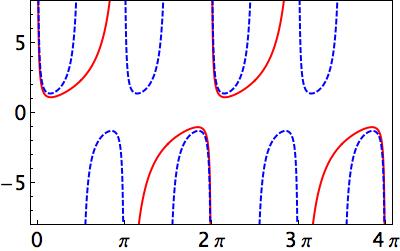} } 

\caption{\footnotesize  The beam width defined in (\ref{width1}) and the corresponding radius of curvature (\ref{R}) for parabolic media, figures~(a) and (b) respectively. In both cases $z_0=0$ and $w_0=1$, with $\Omega =0.5$ (red curve) and $\Omega =1$ (blue-dashed curve). The divergences of $R(z)$ identify the critical values of $w(z)$, and correspond to plane wavefronts.}
\label{Fig1}
\end{figure}

For $\Omega > 0$ the amplitude of the beam width (\ref{width1}) oscillates with period $\pi/\Omega$ between $w_0$ and $w_0/(\Omega z_R)^2$. The latter values are respectively reached at the points $z= \left( n+\tfrac12 \right) \frac{\pi}{\Omega} + z_0$ and $z= n \frac{\pi}{\Omega} + z_0$, with $n=0,1,2,\ldots$ Thus, in contraposition to the homogeneous case where $w(z)$ diverges for large values of $z$ (see Section~\ref{sechom}), the beam width introduced in (\ref{width1}) is finite over all the $z$-axis. Moreover, the maximum amplitude reached by $w(z)$ can be adjusted by varying $\Omega$, see Figure~\ref{Fig1}(a). 

Now let us write explicitly the expression for the radius of curvature. From (\ref{radio}) and (\ref{width1}), one obtains
\begin{equation}
R(z) =\frac{z_R^2 \Omega^2 \cot[\Omega(z-z_0)] + \tan [\Omega (z-z_0)]}{ \Omega( 1 - \Omega^2 z_R^2)}.
\label{R}
\end{equation}
The latter expression diverges at the critical points of $w(z)$, see Figure~\ref{Fig1}(b). Thus, the wavefront of the Gaussian wave-packet (\ref{LG00}) is plane at either the beam waist (minimum beam width) or the beam `hip' (maximum beam width).

\begin{figure}[htb]
\centering
\includegraphics[width=0.25\textwidth]{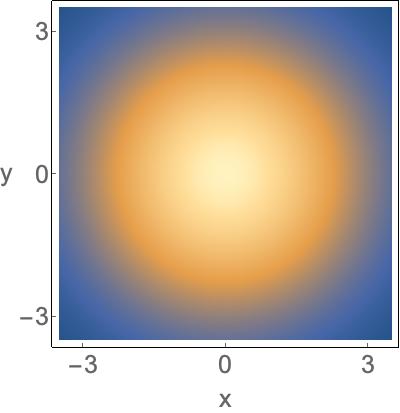} \hskip1ex
\includegraphics[width=0.25\textwidth]{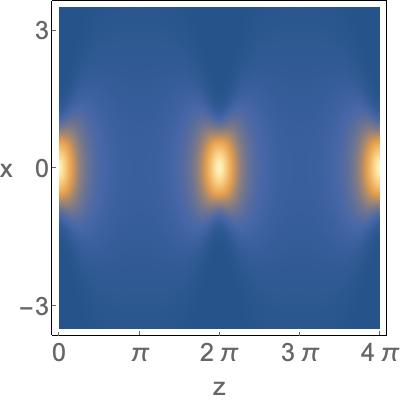}  \hskip1ex
\includegraphics[width=0.3\textwidth]{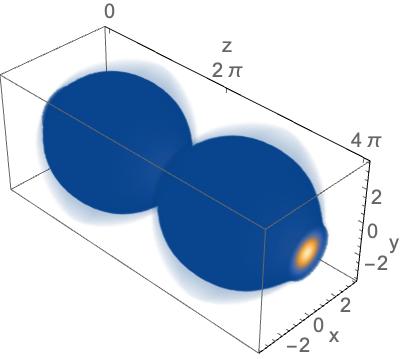}
\vskip1ex
\centering
\includegraphics[width=0.25\textwidth]{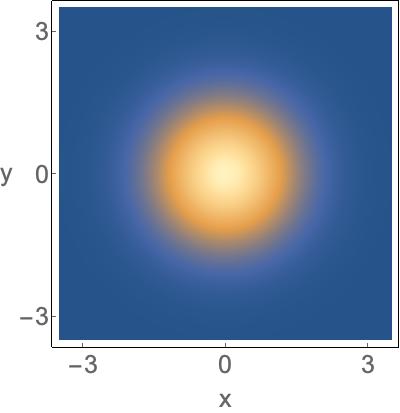} \hskip1ex
\includegraphics[width=0.25\textwidth]{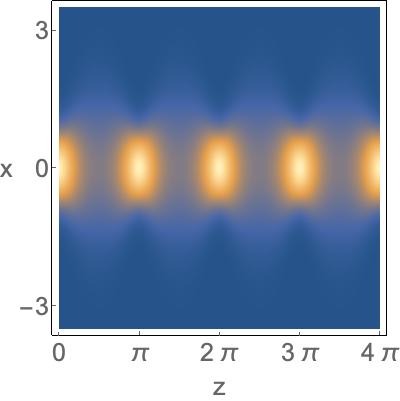} \hskip1ex
\includegraphics[width=0.3\textwidth]{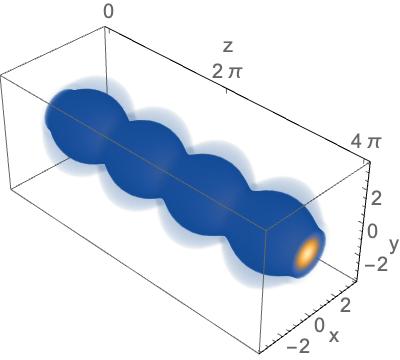}

\caption{\footnotesize Field intensity $\vert U(\rho,z)\vert^2$ of the Gaussian wave-packet (\ref{LG00}) for $z_0=0$ and $k_0=n_0=w_0=1$. The upper and lower rows correspond to $\Omega = 0.5$ and $\Omega =1$, respectively. From left-to-right the columns show the transversal plane where the beam width reaches the first of its maxima, the longitudinal plane $y=0$, and the propagation of the wave-packet along the $z$-axis. Compare with Figure~\ref{Fig1}.}
\label{Fig2}
\end{figure}

On the other hand, the oscillatory profile of the beam width (\ref{width1}) is inherited to the wave-packet (\ref{LG00}). Indeed, the field intensity $\vert U(\rho,z)\vert^2$ describes a spot centered at $\rho=0$ that increases its diameter till a maximum value as the wave-packet propagates along the $z$-axis. Then, the spot starts to reduce its diameter up to recover its initial configuration. This stretching and squeezing phenomenon is repeated over and over  along the propagation axis, see Figure~\ref{Fig2}. Such behavior corresponds to the self-focusing of the beam, which is a consequence of the parabolic profile of the refractive index (\ref{parabolic1}). The number of stretching and squeezing of the beam width in a given interval $z \in (a,b) \subset \mathbb R$ is determined by the parameter $\Omega$, in complete agreement with the periodical profile of the beam width (\ref{width1}). 

\subsubsection{Recovering the results for homogeneous media}
\label{sechom}

\begin{figure}[htb]

\centering
\subfigure[$w_{\operatorname{hom}}(z)$]{\includegraphics[width=0.3\textwidth]{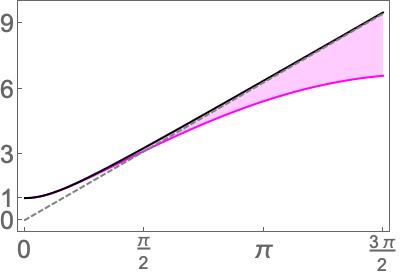} } 
\hskip1cm
\subfigure[$R_{\operatorname{hom}}(z)$]{\includegraphics[width=0.3\textwidth]{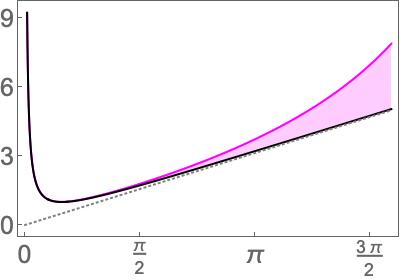} } 

\caption{\footnotesize The beam width $w_{\operatorname{hom}}(z)$ defined in (\ref{wfree}) and the corresponding radius of curvature $R_{\operatorname{hom}}(z)$ for homogenous media, Figures (a) and (b) respectively. In both cases $z_0=0$ and $w_0=1$. The gray-dashed line in (a) identifies the beam angular divergency $\theta_0$. The gray-dotted line in (b) represents the radius of curvature of a spherical wavefront. The magenta curves  respectively correspond to $w(z)$, defined in (\ref{width1}), and to $R(z)$, defined in (\ref{R}), for $\Omega=0.3$. Both of them have been included as a reference. The space between $w_{\operatorname{hom}}(z)$ and $w(z)$, and the one between $R_{\operatorname{hom}}(z)$ and $R(z)$, has been filled by the sake of comparison. See also Figure~\ref{Fig1}.
}
\label{Fig3}
\end{figure}

The description of the propagation in homogeneous media is obtained at the limit $\Omega \rightarrow 0^+$. For instance, at such a limit the beam width acquires the well known form \cite{Sal91,Sie86}:
\begin{equation}
\label{wfree}
w(z)_{\Omega \rightarrow 0^+} = w_{\operatorname{hom}}(z) = w_0 \sqrt{1 + \left(\frac{z-z_0}{z_R}\right)^2}.
\end{equation}
In this case $w_0 = w_{\operatorname{hom}}(z_0)$ corresponds to the beam waist. In turn, $z=z_0$ defines the focal plane and $z_R$ stands for the distance from such plane to the position at which the spot of the beam doubles its size. On the other hand, for large values of $z$, Eq.~(\ref{wfree}) can be approximated as $w_{\operatorname{hom}}(z) \sim \frac{w_0}{z_R} (z-z_0)$. Thus, the beam in homogeneous media diverges as $\vert z \vert \rightarrow \infty$, see Figure~\ref{Fig3}(a), and describes a cone of half-angle defined by the beam angular divergency $\theta_0 \sim \frac{w_0}{z_R}$ (it is subtended by the gray-dashed line with respect to the horizontal axis in the figure). The corresponding radius of curvature $R_{\operatorname{hom}}(z)$ is depicted in Figure~\ref{Fig3}(b), where it is compared with the radius of curvature of a spherical wavefront produced by a point source located at the center of the beam waist (gray-dotted line in the figure), and with the function $R(z)$ defined in (\ref{R}) for $\Omega=0.3$ (magenta curve in the figure). Notice that $R_{\operatorname{hom}}(z)$ has only a singular point, located at $\rho=0$, which means that the wavefront of the corresponding wave-packet is plane at the beam waist only, so no self-focusing is predicted for homogeneous media, as expected.

\subsection{Laguerre-Gaussian wave-packets}

We wonder whether there are other wave-packet solutions to the paraxial wave equation associated with a parabolic refractive index. The answer is positive (see, e.g. \cite{Per96} and \cite{Cru17}) and it may be shown that, using cylindric coordinates, a particularly useful set of solutions can be cast in the form
\begin{equation}
\label{ugen}
U(\boldsymbol{\rho},z) = \frac{\mathcal{N}}{w(z)} 
\exp \left( i \left[\frac{k_0n_0\rho^2}{2R(z)}  - \beta \chi(z)\right]\right) \Phi(r(\rho,z),\theta),
\end{equation}
where $\mathcal N$ and $\beta$ are constants to be determined, $\theta$ stands for the polar coordinate in the transversal plane, and $r(\rho,z) = \sqrt{2} \frac{\rho}{w(z)}$ is the corresponding radial coordinate. The straightforward calculation shows that the function $\Phi(r,\theta)$ satisfies the differential equation
\begin{equation}
\label{twoosc}
- \nabla_{\perp}^2\Phi + r^2 \Phi = 2 \beta \Phi,
\end{equation}
which resembles the stationary Schr\"odinger equation for a two dimensional oscillator in the radial variable $r$. The resemblance is complete if one considers that square integrability in the Hilbert space corresponds to finite transverse optical power $P_0$ for localized beams in the $(r,\theta)$-plane. That is, we demand the field intensity of the wave-packets (\ref{ugen}) to satisfy the condition
\[
P_0 = \int_0^{2\pi} \int_0^\infty \left|U_\ell^p(\boldsymbol{\rho},z)\right|^2 \rho d\rho d\theta = 1.
\]

\begin{figure}
\centering
\includegraphics[width=0.25\textwidth]{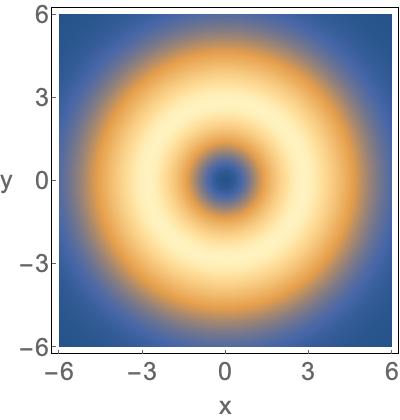} \hskip1ex
\includegraphics[width=0.25\textwidth]{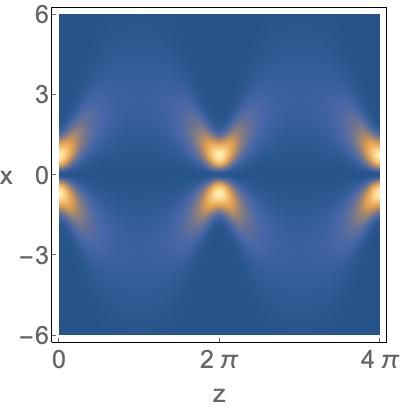}  \hskip1ex
\includegraphics[width=0.3\textwidth]{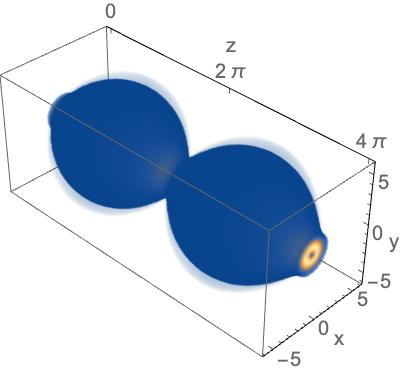}
\vskip1ex
\includegraphics[width=0.25\textwidth]{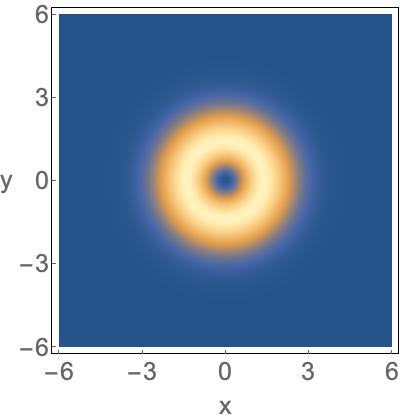} \hskip1ex
\includegraphics[width=0.25\textwidth]{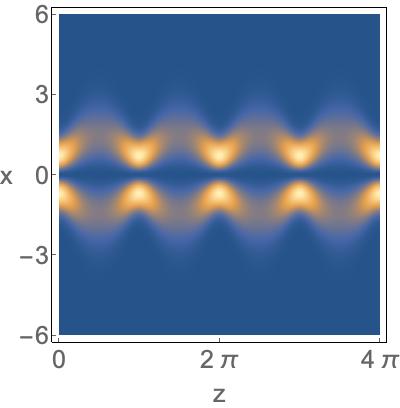} \hskip1ex
\includegraphics[width=0.3\textwidth]{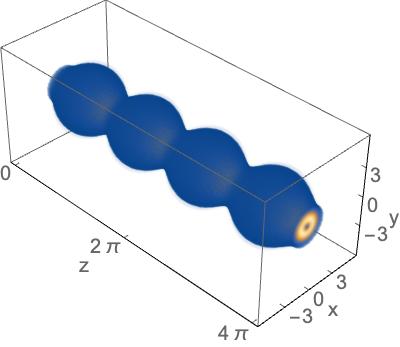}

\caption{\footnotesize Field intensity $\vert U_{\ell}^p (\boldsymbol{\rho}, z) \vert^2$ of the Laguerre-Gaussian wave-packet (\ref{ugen}) for $p=0$ and $\ell=1$, with the same parameters and distribution as the panel shown in Figure~\ref{Fig2}.
}
\label{Fig4}
\end{figure}

The conventional approach used to face stationary problems in quantum mechanics yields the eigenvalues
\begin{equation}
\beta_\ell^p = \vert \ell \vert + 2p + 1, \quad \ell \in \mathbb Z, \quad p=0,1,2,\ldots,
\label{eigen1}
\end{equation}
together with the eigenfunctions
\begin{equation}
\Phi_{\ell}^p (r,\theta) = \left(\frac{\sqrt{2}\rho}{w(z)}\right)^{\vert \ell \vert}L_p^{(\vert \ell \vert)} \left(\frac{2\rho^2}{w^2(z)}\right)  \exp \left(-\frac{\rho^2}{w^2(z)} + i \ell \theta \right), 
\label{phi}
\end{equation}
and the constant
\begin{equation}
\mathcal{N}_{\ell}^p = (-1)^p \sqrt{\frac{2 \Gamma(p + 1)}{\pi \Gamma(\vert \ell \vert + p + 1)}}.
\label{Ncons}
\end{equation}
The introduction of (\ref{eigen1})-(\ref{Ncons}) into (\ref{ugen}) gives the expressions $U_{\ell}^p(\boldsymbol{\rho},z)$ we were looking for. Hereafter they are referred to as Laguerre-Gaussian (LG) wave-packets.

\begin{figure}
\begin{center}
\includegraphics[width=0.25\textwidth]{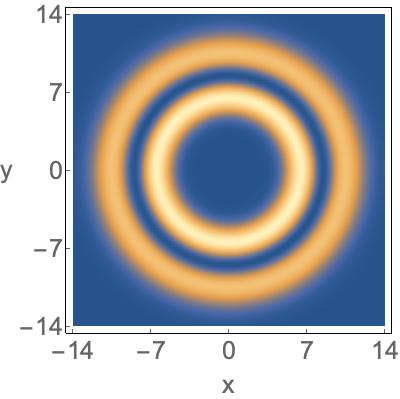} \hskip1ex
\includegraphics[width=0.25\textwidth]{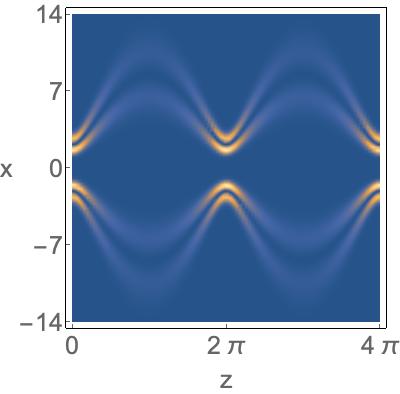}  \hskip1ex
\includegraphics[width=0.3\textwidth]{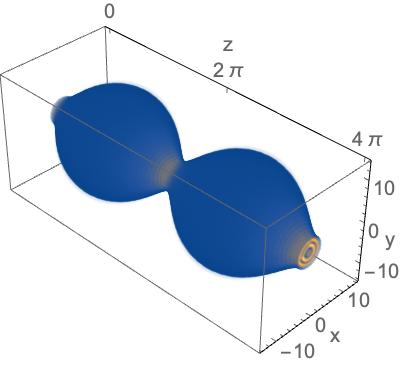}
\vskip1ex
\includegraphics[width=0.25\textwidth]{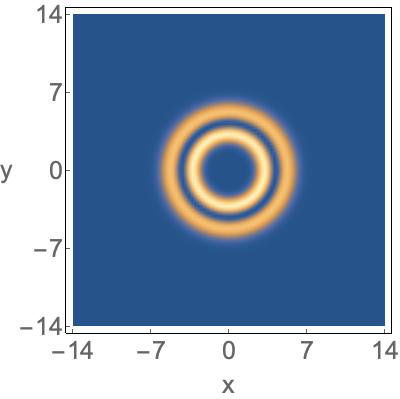} \hskip1ex
\includegraphics[width=0.25\textwidth]{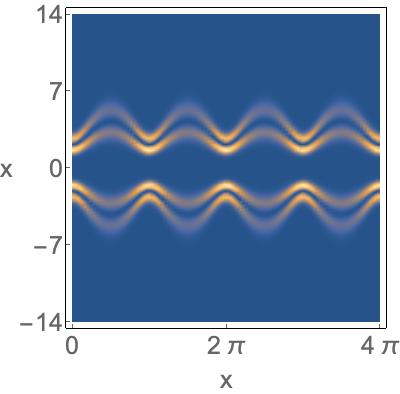}  \hskip1ex
\includegraphics[width=0.3\textwidth]{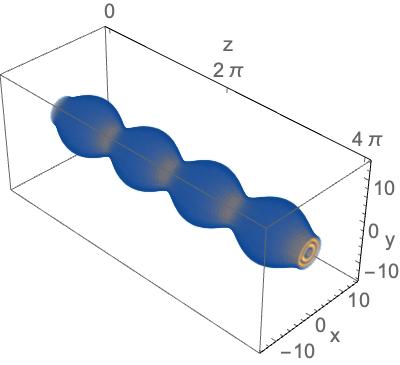}

\caption{\footnotesize Field intensity $\vert U_{\ell}^p (\boldsymbol{\rho}, z) \vert^2$ of the Laguerre-Gaussian wave-packet (\ref{ugen}) for $p=1$ and $\ell=8$, with the same parameters and distribution as the panel shown in Figure~\ref{Fig2}.
}
\label{Fig5}
\end{center}
\end{figure}

The results discussed in Section~\ref{lowest} for the nonstationary Gaussian wave-packets are recovered from the above formulae after making $p=0$ and $\ell =0$, which respectively means the lowest radial parameter and null orbital angular momentum. For $p=0$ and $\ell \neq 0$ the field intensity $\vert U_{\ell}^p (\boldsymbol{\rho}, z) \vert^2$ describes a ring centered at $\rho=0$, see Figure~\ref{Fig4}. In general, for $\ell \neq 0$ and any allowed value of $p$, the field intensity of the LG wave-packets (\ref{ugen}) exhibits a well known ring-shaped distribution. The size of the rings depends on $\ell$ while $p$ defines the number of nodes in the radial coordinate, see Figure~\ref{Fig5} and compare with Figs.~\ref{Fig2} and \ref{Fig4}. Thus, controling the values of $p$, $\ell$, and $\Omega$, one can manipulate the profile as well as the collapse-revival (self-focusing) properties of the LG wave-packets (\ref{ugen}). 

\subsubsection{Recovering the results for guided Laguerre-Gaussian modes}

If the parameters $w_0$ and $\Omega$ are related as $k_0n_0 \Omega = 2/w_0 ^2$, then $\Omega z_R = 1$ and, according to (\ref{width1}),  the beam width becomes a constant $w(z)=w_0$, see \cite{Gre19}. In such a case the radius of curvature $R(z)$ diverges and the Gouy phase turns into a linear function of $z$. Thus, the wavefront of the LG wave-packets is plane. The corresponding field amplitudes, known as guided LG modes, take the form
\begin{equation}
\begin{array}{rl}
U_\ell^p(\boldsymbol{\rho},z)  & = (-1)^p \sqrt{\frac{2 \Gamma(p + 1)}{\pi w_0^2\Gamma(\vert \ell \vert + p + 1)}}  \left(\frac{\sqrt{2}\rho}{w_0}\right)^{\vert \ell \vert}L_p^{(\vert \ell \vert)} \left(\frac{2\rho^2}{w^2_0}\right) \\[2ex]
& \qquad \times  \exp \left( -\frac{\rho^2}{w^2_0} -\beta_\ell^p \Omega (z-z_0) + i \ell \theta \right),
\end{array}
\end{equation}
and are stationary eigenmodes of the Schr\"odinger-like operator 
\begin{equation}
H= -\frac1{2 k_0^2 n_0} \nabla_{\perp}^2  + \frac{n_0}2\Omega^2 \rho^2,
\end{equation}
with the propagation constants
\begin{equation}
\varepsilon_\ell^p = \frac{\Omega}{k_0} \beta_\ell^p  \equiv \frac{\Omega}{k_0}(\vert \ell \vert + 2p + 1), \quad \ell \in \mathbb Z, \quad p=0,1,2,\ldots
\end{equation}


\section{Discussion of results}
\label{discute}

We have shown that the beam width of the Laguerre-Gaussian wave-packets is finite and periodic along all the propagation axis if it is a solution of the Ermakov equation associated with the one-dimensional harmonic oscillator. The amplitude of the beam width can be modulated by the strength of the medium inhomogeneity. The wave-packets so constructed have finite transverse optical power and propagate with periodical self-focusing in the medium. The conventional Laguerre-Gaussian modes are recovered as a particular case, after turning-off the inhomogeneity. 

Since the orbital angular momentum is a consequence of the mode structure of a given beam, one would guess that such a property is also present in single photons \cite{Cha18}. The statement seems to be strengthened by recalling that idealized plane waves with only transverse fields do not carry angular momentum, no matter their degree of polarization \cite{Sim70}. A clue may be found in the phenomenon of parametric-down conversion \cite{Pro10,Pro15}, where the quantum state of down-converted photon pairs is entangled in at least one of their physical variables. As entanglement is a fingerprint of the quantum world, the production of photon pairs entangled in their orbital angular momentum states \cite{Fra02} bets on a quantum nature of structured light. It is then interesting to formulate an approach based on operators in Hilbert spaces to describe the propagation of structured light beams in diverse media. Some initial steps on the matter were given in \cite{Enk92}. The complete operator-like description for the propagation of Hermite-Gaussian wave-packets in parabolic media has been provided in \cite{Cru17,Gre17,Gre19}, where the Lie group formalism was addressed to construct generalized coherent states as linear superpositions of Hermite-Gaussian modes. The same approach can be applied to the Laguerre-Gaussian wave-packets reported in this paper. Work in this direction is in progress.


\subsection*{Acknowledgment}
This research was funded by Consejo Nacional de Ciencia y Tecnolog\'ia (Mexico), grant number A1-S-24569, and by Instituto Polit\'ecnico Nacional (Mexico), project SIP20195981.


\end{document}